 \documentclass[preprint]{aastex61}

\begin{document}

\title{The acceleration of high-energy protons at coronal shocks: the effect of large-scale streamer-like magnetic field structures}

%%\correspondingauthor{Xiangliang Kong}
%%\email{kongx@sdu.edu.cn}

\author{Xiangliang Kong}
\affil{Shandong Provincial Key Laboratory of Optical Astronomy and Solar-Terrestrial Environment,
and Institute of Space Sciences, Shandong University, Weihai, Shandong 264209, China; kongx@sdu.edu.cn}
\affiliation{Los Alamos National Laboratory, Los Alamos, NM 87545, USA; guofan.ustc@gmail.com}

\author{Fan Guo}
\affiliation{Los Alamos National Laboratory, Los Alamos, NM 87545, USA; guofan.ustc@gmail.com}
\affiliation{New Mexico Consortium, Los Alamos, NM 87544, USA;}

\author{Joe Giacalone}
\affiliation{Department of Planetary Sciences, University of Arizona, Tucson, AZ 85721, USA}

\author{Hui Li}
\affiliation{Los Alamos National Laboratory, Los Alamos, NM 87545, USA; guofan.ustc@gmail.com}

\author{Yao Chen}
\affil{Shandong Provincial Key Laboratory of Optical Astronomy and Solar-Terrestrial Environment,
and Institute of Space Sciences, Shandong University, Weihai, Shandong 264209, China; kongx@sdu.edu.cn}

\begin{abstract}

Recent observations have shown that coronal shocks driven by coronal mass ejections can develop and accelerate particles within several solar radii in large solar energetic particle (SEP) events. 
Motivated by this, we present an SEP acceleration study including the process that a fast shock propagates through a streamer-like magnetic field with both closed and open field lines in the low corona region.
The acceleration of protons is modeled by numerically solving the Parker transport equation with spatial diffusion both along and across the magnetic field. 
We show that particles can be sufficiently accelerated to up to several hundred MeV within 2-3 solar radii. 
When the shock propagates through a streamer-like magnetic field, particles are more efficiently accelerated compared to the case with a simple radial magnetic field, mainly due to perpendicular shock geometry and the natural trapping effect of closed magnetic fields.
Our results suggest that the coronal magnetic field configuration is an important factor for producing large SEP events. 
We further show that the coronal magnetic field configuration strongly influences the
distribution of energetic particles, leading to different locations of source regions along the shock front where most of high energy particles are concentrated. 
This work may have strong implications to SEP observations. 
The upcoming Parker Solar Probe will provide in situ observations for the distribution of energetic particles in the coronal shock region, and test the results of the study.

\end{abstract}


\section{Introduction} \label{sec:intro}

Solar energetic particles (SEPs) are accelerated near the Sun during explosive solar activities such as flares and coronal mass ejections (CMEs) \citep{reames99,desai16}. 
Large SEP events are of particular importance because they pose severe radiation hazards to astronauts and equipment in space. 
Those events are featured by the intense proton flux in the $>$10 MeV GOES energy channel (exceeding 10 particle cm$^{-2}$ s$^{-1}$ sr$^{-1}$, or particle flux unit) and there have been about 10 events observed per year on average during solar maximum. 
Ground level enhancements (GLEs) are a type of extremely large SEP events, in which the highest-energy ($\sim$GeV) particles can produce secondaries detectable by ground-based instruments when those particles interact with Earth's atmosphere. 
While 16 GLE events were recorded in solar cycle 23, there have been only 2 GLEs observed so far in solar cycle 24 \citep{gopalswamy13,thakur14}.
Meanwhile, the number of large SEP events in this solar cycle is significantly less than that of the last solar cycle \citep[e.g.,][]{mewaldt15,richardson17}. 
It is known that the solar activity is weak in the current solar cycle, however, the physical reason for the lack of large SEP events and GLEs is still not clear \citep[see, e.g.,][]{gopalswamy13,gopalswamy14,giacalone15,mewaldt15}.

In large SEP events, high energy particles are believed to be accelerated by CME-driven shock waves \citep{reames99,desai16}. 
There have been a number of factors regarding the CME-shock properties found to be important in producing large SEP events.
For example, large SEP events are often associated with fast and very energetic CMEs \citep[e.g.,][]{kahler01,mewaldt08}.
Especially, the CMEs that produce GLE events are among the fastest ones with speeds of $\sim$2000 km s$^{-1}$ \citep[e.g.,][]{gopalswamy12,gopalswamy13,thakur14}. 
Those CMEs are usually in the western hemisphere and have a good magnetic connectivity to the Earth \citep[e.g.,][]{gopalswamy12,gopalswamy14}. 
However, not all CMEs that satisfy the two factors can lead to large SEP events or GLEs \citep[e.g.,][]{kahler05,ding13}, indicating other processes are important in producing large SEP events as well.
The proposed factors include the background energetic particles that provide seed particles for the shock acceleration to high energies \citep[e.g.,][]{kahler00,kahler01,cliver06}, preceding CMEs that may provide both seed particles and enhanced turbulence for more efficient particle acceleration \citep[e.g.,][]{gopalswamy04,li12,ding13,zhao14}, 
and the CME-shock geometry that allows perpendicular shock region to rapidly accelerate particles to high energies \citep[e.g.,][]{tylka05}.

An important piece of information on the acceleration of SEPs is about the source regions.
Recent observations have provided convincing evidence that shocks can form and accelerate particles low in the corona.
\citet{ma11} suggested that the brightness enhancement ahead of the bright CME front seen by extreme ultraviolet observations of SDO/AIA is a shock signature and it forms as low as $\sim$1.2 $R_\odot$ from the center of the Sun. 
\citet{gopalswamy13b} found that CME-shock formation can occur substantially below $\sim$1.5 $R_\odot$ from observations of type II radio bursts produced by electrons accelerated by shocks.
\citet{gopalswamy12} analyzed GLEs in solar cycle 23 and found that while the shock formation height can be $\sim$1.5 $R_\odot$, 
the CME heights at GLE particle release times range from 1.7 to 4 $R_\odot$, with an average of 3.1 $R_\odot$ for well-connected events, suggesting that particle acceleration occurs in this region.
The results are in line with earlier studies for GLEs in both solar cycle 23 and previous cycles using the velocity dispersion analysis by \citet{reames09a,reames09b}. 
Consistent results have also been obtained for the two GLEs in solar cycle 24 \citep{gopalswamy13,thakur14}.
Furthermore, \citet{gopalswamy16} found that the CME initial speed and acceleration are larger in GLEs compared to regular SEP events and the SEP spectral indices are related to the CME initial acceleration, also indicating that it is important to have a shock that accelerates particles low in the corona to produce large SEP events.
The pieces of evidence above suggest that a CME-shock is likely to form and start accelerating particles very close to the Sun (well below 2 $R_\odot$).

The primary theory of particle acceleration at shocks is diffusive shock acceleration (DSA) \citep{axford77,krymsky77,bell78,blandford78}. 
In the presence of magnetic turbulence, particles can be scattered back and forth across the shock multiple times and get efficient acceleration.
The DSA naturally predicts a power-law energy spectrum with the spectral index only depending on the shock compression ratio.
The acceleration of particles may significantly depend on the angle $\theta_{Bn}$ between the upstream magnetic field and the shock normal.
Particle acceleration at a nearly perpendicular shock is much faster than that at a parallel shock \citep{jokipii87}. 
Note that, however, self-excited waves produced by streaming energetic particles may further enhance particle acceleration at parallel shocks \citep[e.g.,][]{lee83,lee05,li03,rice03}. 
When a CME-driven shock develops in the corona, the nonplanar shock front sweeps through the coronal magnetic field with a range of different shock angles, 
which has been proposed to have significant effects on particle acceleration \citep[e.g.,][]{giacalone05a,giacalone05b,tylka05}.
By including large-scale magnetic field variation, the transport effect is found to be important in the shock region, creating hot spots where most of high energy particles are concentrated \citep{guo10}.

When a CME-driven shock propagates outwards from the corona to the interplanetary space, the acceleration rate can vary considerably. 
The strong magnetic field close to the Sun likely gives rise to a small diffusion coefficient and thus more efficient particle acceleration based on the DSA.
In addition, a CME-shock itself may weaken as it propagates away from the Sun.  Therefore, the maximum achievable particle energy decreases with heliocentric distance \citep[e.g.,][]{zank00,li05b}.
In particular, as mentioned earlier, compelling observational evidence exists for the shock formation and particle acceleration close to the Sun. 
Therefore, modeling shock acceleration in the low corona is of great importance for understanding the underlying mechanisms for producing large SEP events
\citep[e.g.,][]{berezhko03,berezhko13,kocharov05,vainio07,ng08,sandroos09a,sandroos09b,kozarev13,schwadron15}. 
\citet{ng08} modeled particle acceleration at a parallel shock with a speed of 2500 km s$^{-1}$ starting at 3.5 $R_\odot$ and showed that particles can be accelerated to $>$300 MeV in 10 minutes.
\citet{sandroos09a} studied the effect of evolving shock geometry on the acceleration efficiency as a coronal shock moves through a large-scale radial coronal magnetic field and found that for most considered field lines the maximum proton energies are $>$100 MeV but the intensities can vary by more than two orders of magnitude. 
A key issue that is not well known is the primary acceleration region along the CME shock where most of high energy particles are produced. While several studies have shown that the large gradual SEP events can be understood under the assumption that the shock nose is the main source of energetic particles \citep[e.g.,][]{cane88,reames97,reames99}, some other studies have suggested that the shock flank can provide the majority of energetic particles \citep[e.g.,][]{kahler16}.
We note here that the role of magnetic field configuration is of great importance in determining the primary acceleration region.

In this work, we emphasize the importance of large-scale streamer-like magnetic fields to particle acceleration at coronal shocks.
As seen from coronagraph images, streamers are the most prominent quasi-steady structures in the corona and CMEs often closely interact with streamers as they propagate outwards or expand laterally \citep[e.g.,][]{chen10,feng11}.
Statistical analyses have indicated that many CMEs originate from the streamer belt
\citep[e.g.,][]{howard85,hundhausen93,eselevich97,zhao03}.
Furthermore, recent works have shown that type II radio bursts, which are good observational manifestations of energetic electrons accelerated at shocks, reflect the interaction between shocks and streamers
\citep[e.g.,][]{reiner03,mancuso04,cho08,feng12,feng13,kong12,kong15,chen14} 
or the shock passing through high-density coronal loops 
\citep[e.g.,][]{pohjolainen08,cho13}.
Using test-particle simulations, \citet{kong15,kong16} studied the effect of streamer-like magnetic fields on electron acceleration at coronal shocks and found that the closed field lines can trap a significant amount of electrons close to the shock front and lead to an efficient acceleration.

Lately, \citet{rouillard16} showed that the CME-shock of the GLE event on 2012 May 17 originated below a streamer, indicating that the large-scale magnetic configuration of streamers may also affect SEP acceleration at coronal shocks.
However, due to limited capability of current imaging observations of CME-streamer interaction in the corona and unavailability of direct measurement of coronal magnetic field topology, the role of streamers or CME-streamer interaction in the acceleration, trapping and release of energetic particles remains unclear \citep[see, e.g.,][]{kahler00,kahler05,rouillard16,kocharov17}.  
Here we present a numerical model to investigate particle acceleration at shocks close to the Sun, with a focus on the possible role of large-scale streamer-like magnetic configuration in producing large SEP events.
We consider a coronal shock with a kinematic description propagating through a streamer-like magnetic field that is analytically given \citep{low86}.
Particle acceleration is modeled by numerically solving the Parker transport equation \citep{parker65} with spatial diffusion both along and across the magnetic field.
The variations of particle spectra and spatial distribution with the shock propagation are investigated.

The rest of this paper is organized as follows: 
In Section 2, we describe the numerical model for solving the energetic particle transport equation in a streamer structure involving both closed and open magnetic fields.
We present the simulation results including the energy spectra and spatial distribution of energetic particles in Section 3. We compare the results for the streamer structure with that for a radial magnetic field case. We also present analysis on the acceleration process of high energy particles. 
Our conclusions and discussion are given in Section 4.

\section{Numerical Model} \label{sec:model}
\begin{figure}%[\htb]
%% trim: left, bottom, right, top
\includegraphics[width=0.7\textwidth,clip,trim=0cm 1cm 5cm 3cm]{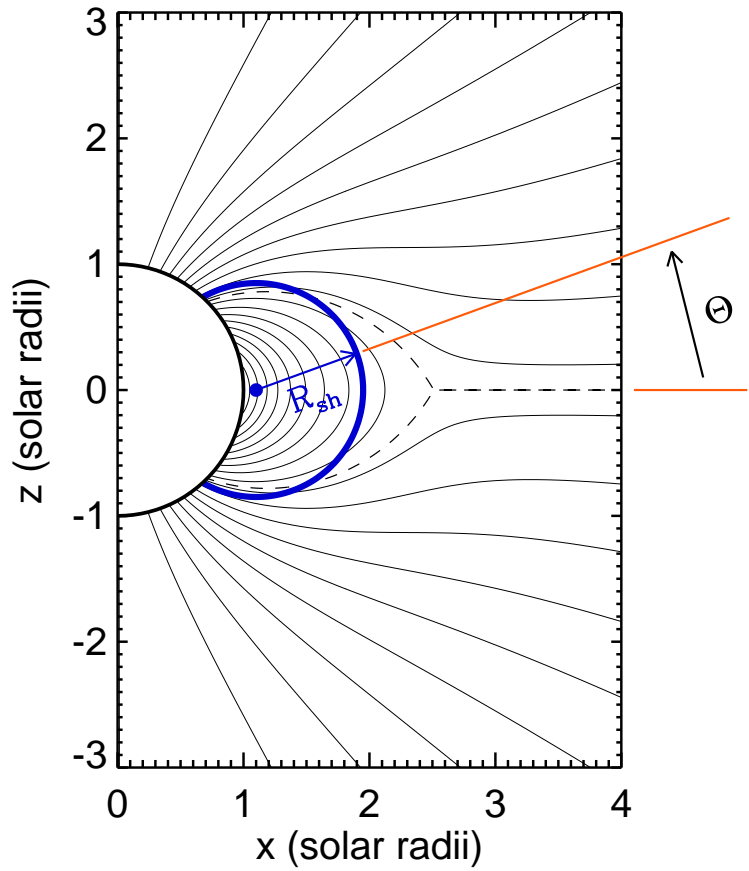}
\caption{Schematics of the coronal shock morphology (thick blue circle) and the streamer-like coronal magnetic field (black lines). The blue dot indicates the center of the shock, fixed at 0.1 $R_\odot$ above the solar surface.
}\label{fig:Fig1}
\end{figure}

In our model, we consider proton acceleration at a fast coronal shock propagating through a streamer structure with both closed and open magnetic field lines. 
A schematic of the numerical model is shown in Figure \ref{fig:Fig1}.
The traveling coronal shock is represented by an expanding circular front. 
The coronal magnetic field is in a partially open streamer-like configuration \citep{low86}. 
For comparison, we also consider a case with only radial magnetic field from the center of the Sun. 
We study the acceleration of protons in the shock-streamer system by numerically solving the Parker transport equation \citep{parker65} through a stochastic integration method \citep[e.g.,][]{giacalone08,guo10,senanayake13}.

We use two coordinate systems simultaneously. 
The helicocentric coordinate system ($x$, $y$, $z$) has its origin at the center of the Sun, with the $z$-axis being the solar rotation axis, the $x$-axis towards the intersection of solar equator and solar central meridian, and the $y$-axis completing the right-handed triad. 
% Helicocentric Earth Equatorial (HEEQ), as seen from Earth
In this study, both the coronal shock and the background magnetic field are assumed to be axisymmetric about the $z$-axis, and we only consider a two-dimensional simulation in the $x$-$z$ plane. 
The shock coordinate system in the $x$-$z$ plane ($R$, $\Theta$) has its origin at the center of the shock. The $R$ direction points outward radially from the shock center (i.e., along the shock normal), and $\Theta$ is defined as the angle with respect to the streamer axis.

\subsection{Shock and magnetic fields}

As shown in Figure \ref{fig:Fig1}, the shock center is located in the solar equatorial plane at a height $h =$ 0.1 $R_\odot$ ($x=1.1$ $R_\odot$, $z=0$) above the solar surface and is fixed for simplicity.
It can be a reasonable assumption in the low corona since the CME-shock expansion is usually dominant in the initial phase of CME eruptions \citep[e.g.,][]{kwon14}.
The circular shock front propagates outward with a constant expansion speed $V_{sh}$ = 2000 km s$^{-1}$ and a compression ratio $X$ = 3. 
The shock radius $R_{sh} = R_{sh0} + V_{sh}t$, where $R_{sh0}$ = 0.2 $R_\odot$ is taken to be the initial radius of the shock.

We assume that the background solar wind speed is much less than the shock speed and set it to zero.
Using a non-zero solar wind speed will lead to a slightly weaker shock.
In the local shock frame of reference, the fluid velocity along the shock normal is approximated by a hyperbolic tangent function,
\begin{equation} 
 U(x') = \frac{U_1 + U_2}{2} - \frac{U_1 - U_2}{2} tanh \left(\frac{x'}{\delta_{sh}} \right),
 \end{equation} 
where $x'$ is the distance to the shock front along the shock normal, $U_1$ = $V_{sh}$ and $U_2$ = $V_{sh}$/$X$ are the upstream and downstream normal flow speeds, and $\delta_{sh}$ is the width of the shock. 
In the laboratory frame, the plasma flows radially from the shock center, with the velocity in the shock upstream being zero, i.e., $V_1 = 0$, 
in the downstream $V_2 = V_{sh}(1-1/X)$,
and in the shock layer $V = V_{sh} - U$.

Following our previous studies \citep{kong15,kong16}, the background coronal magnetic field is taken to be an analytical solution of a streamer-like configuration \citep{low86}.
A brief introduction to the analytical solution is presented in the Appendix.
It describes an axisymmetric magnetic structure containing both closed magnetic arcades and open field lines with a current sheet.
%\textcolor{red}{in a spherical coordinate (remove). BTW: Are you sure it is a spherical coordinate? (yes)}
This coronal magnetic field model has been used in previous corona and solar wind modelings \citep[e.g.,][]{chen01,hu03}. 
In this study, %we only consider a two-dimensional simulation in the $x$-$z$ plane and 
the azimuthal component of magnetic field is set to be zero.
The height of the streamer cusp is taken to be 2.5 $R_\odot$.
The magnetic field strength on the solar surface is set to be $\sim$5 G at the equator and $\sim$10 G in polar regions.
For comparison, we also use a large-scale radial magnetic field with the strength decreasing as a function of $B = 5(R_\odot/r)^2$ G, where $r$ is the heliocentric distance.
%\textcolor{red}{Note:in 2D, the decay of magnetic field should be 1/R. }

The magnetic field is compressed when it is swept by the shock into the downstream region. 
In the shock layer (within 8 $\delta_{sh}$ from the shock front), the magnetic field is determined by the MHD shock jump conditions in the local shock frame. 
The tangent component of magnetic field with respect to the shock surface is given by $B_{\Theta} = B_{\Theta 0} V_{sh}/U$, where $B_{\Theta 0}$ is the tangent component of background field, while the normal component $B_R$ remains unchanged.
In the downstream region, %(with a distance more than 8 $\delta_{sh}$ from the shock front), 
the magnetic field $\textbf{B}_2$ is obtained from the induction equation,
%$\textbf{E}_2 = - \textbf{V}_2 \times \textbf{B}_2$,
\begin{equation} 
\frac{\partial \textbf{B}_2}{\partial t} = \nabla \times (\textbf{V}_2 \times \textbf{B}_2).
\end{equation} 
Here $\textbf{V}_2$ is the vector of downstream velocity in the direction of shock normal. 
The induction equation can be written as first-order linear partial differential equations of the downstream magnetic field components $B_{R2}$ and $B_{\Theta 2}$, and solved analytically by using the values immediately downstream of the shock at time $t' = t - (R_{sh} - R)/U_2$ as the initial condition.
%The geometry of magnetic field lines after the shock compression has been shown in Figure \ref{fig:Fig1}.
%(will try to plot it later, using contour of Az)

\subsection{Particle acceleration}

We study the acceleration of particles by  numerically integrating the Parker transport equation \citep{parker65},
\begin{equation}  % \[
\frac{\partial f}{\partial t} = 
\frac{\partial}{\partial x_i} \left[ \kappa_{ij} \frac{\partial f}{\partial x_j} \right] 
- U_i \frac{\partial f}{\partial x_i}
+ \frac{p}{3} \frac{\partial U_i}{\partial x_i} \frac{\partial f}{\partial p} 
+ Q,
\end{equation} % \]
where $f(x_i, p, t)$ is the particle distribution function as a function of the particle position $x_i$,  momentum $p$, and time $t$;
$\kappa_{ij}$ is the spatial diffusion coefficient tensor, $U_i$ is the bulk plasma velocity, and $Q$ is the source. 
Note that the gradient and curvature drifts are in the out of the $x$-$z$ plane direction and irrelevant to this study.
The effect of adiabatic cooling of particles due to non-zero flow divergence in the downstream region is included in the calculation.

The diffusion coefficient tensor is given by,
\begin{equation} 
\kappa_{ij} = \kappa_{\perp} \delta_{ij} + (\kappa_{\parallel} - \kappa_{\perp}) \frac{B_i B_j}{B^2},
\end{equation} 
where $\kappa_{\parallel}$ and $\kappa_{\perp}$ are the diffusion coefficients parallel and perpendicular to the large-scale magnetic field. We assume a well-developed magnetic turbulence with a Kolmogorov power spectrum $P \propto k^{-5/3}$ in the coronal region. 
$\kappa_{\parallel}$ can be calculated from the quasilinear theory  \citep{jokipii71}. 
The resulting diffusion coefficient has a momentum dependence, $\kappa_{\parallel} \propto p^{4/3}$, when the particle gyroradius is much smaller than the correlation length of turbulence.
Test-particle calculation has suggested that $\kappa_{\perp}$ is approximately a few percent (0.02 - 0.04) of $\kappa_{\parallel}$ and 
their ratio $\kappa_{\perp} / \kappa_\parallel$ is nearly independent of particle energy  \citep{giacalone99}.
We therefore take the form $\kappa_\parallel = \kappa_{\parallel 0} (p/p_0)^{4/3}$ and $\kappa_{\perp}$ = 0.04 $\kappa_\parallel$, using a smaller $\kappa_{\perp}$ will result in a higher maximum energy. 
The initial momentum $p_0$ corresponds to the energy $E_0$ = 100 keV. 
The value of $\kappa_{\parallel}$ has the following expression \citep{giacalone99},
\begin{equation} 
\kappa_{\parallel} = \frac{3 v^3}{20 L_c \Omega^2 \sigma^2} 
csc\left(\frac{3 \pi}{5}\right) \left[1 + \frac{72}{7} \left(\frac{\Omega L_c}{v} \right)^{5/3} \right],
\end{equation}  %}
where $v$ is the particle velocity, $L_c$ is the turbulence correlation length, $\sigma^2$ is the normalized wave variance of turbulence, and $\Omega$ is the particle gyro-frequency.
We assume the average magnetic field $B_0$ = 1 G, the turbulence amplitude $\sigma^2 = \delta B^2/B_0^2$ = 1, and the turbulence correlation length $L_c$ = 0.2 $R_\odot$, which gives $\kappa_{\parallel 0} = 1.4 \times 10^{17}$ cm$^2$ s$^{-1}$ for the proton initial energy $E_0$ = 100 keV. 
This value is similar to the choice of previous work \citep{Sokolov2004,kozarev13}. Future work will study the influence of the diffusion coefficients to the results of the paper.

Strictly speaking, when we consider particle scattering by magnetic fluctuations the wave propagation needs to be taken into account as the scattering centers move away from the shock. Considering this leads to a steeper energy spectrum and lower energy gain for each shock crossing, but the effect is small if the shock is sufficiently strong \citep{bell78}.
%(0.1, 0.3, 0.5)
% (0.005, 0.03, 0.06)

In the simulation, the normalized parameters are: the length  $L_0$ = 1 $R_\odot$ = 7 $\times 10^{5}$ km, the velocity $V_0$  = $V_{sh}$ = 2000 km s$^{-1}$, the diffusion coefficient $\kappa _0$ = $L_0 V_0$ = 1.4 $\times 10^{19}$ cm $^2$ s$^{-1}$, and the time $t_0 = L_0/V_0 =$ 350 s.

We use a stochastic integration method to obtain the numerical solution of the Parker transport equation (Equation 3).
In this approach, the transport equation can be written in the form of a Fokker-Planck equation and the solution can be calculated by successively integrating the trajectories of pseudo-particles using stochastic differential equations \citep[e.g.,][]{guo10}:  
\begin{equation} 
\Delta x = r_1 \sqrt{2 \kappa_{\perp} \Delta t}  + r_3 \sqrt{2 (\kappa_{\parallel} - \kappa_{\perp}) \Delta t} \frac{B_x}{B} + U_x \Delta t 
         +  \left( \frac{\partial \kappa_{xx}}{\partial x} + \frac{\partial \kappa_{xz}}{\partial z}  \right) \Delta t, 
\end{equation}
\begin{equation} 
\Delta z = r_2 \sqrt{2 \kappa_{\perp} \Delta t}  + r_3 \sqrt{2 (\kappa_{\parallel} - \kappa_{\perp}) \Delta t} \frac{B_z}{B} + U_z \Delta t 
         +  \left( \frac{\partial \kappa_{zz}}{\partial z} + \frac{\partial \kappa_{xz}}{\partial x}  \right) \Delta t,
\end{equation}
\begin{equation} 
\Delta p = - \frac{p}{3} \left( \frac{\partial U_x}{\partial x} + \frac{\partial U_z}{\partial z} \right) \Delta t,
\end{equation}
where $r_1$, $r_2$, and $r_3$ are different sets of random numbers which satisfy $\langle r_i \rangle$ = 0 and $\langle r_i^2 \rangle$ = 1.

%\textcolor{red}{At each time step $\Delta t$, the spatial change $\Delta x$ mainly contains two parts: the convective step $\Delta x_c$ = $V_0  \Delta t$ 
%and the diffusive step $\Delta x_{d} = \sqrt{2 \kappa_{xx} \Delta t}$.
%The characteristic diffusion length scale is $L_{d} = \kappa_{xx} /V_0$.
%For a planar shock, $\kappa_{xx}$ is the total diffusion coefficient normal to the shock front at an oblique shock angle $\theta_{Bn}$, given by $\kappa_{xx} = \kappa_{\perp} sin^2\theta_{Bn} + \kappa_{\parallel} cos^2\theta_{Bn}$. }

%\textcolor{red}{
To get the correct solution of DSA, the following conditions should be satisfied. 
First, the shock thickness $\delta_{sh}$ should be much smaller in comparison to the characteristic diffusion length at the shock front $L_{d} = \kappa_{nn} /V_{sh}$, where $\kappa_{nn}$ is the diffusion coefficient in the direction normal to the shock. 
We take the value $\delta_{sh}$ = 4 $\times$ 10$^{-5}$ to ensure that this is satisfied throughout the simulation. Second, the time step needs to be small enough for particles to ``see" the shock transition. We take the value $\Delta t = 3.2 \times 10^{-7}$ at the initial momentum and the time step is reduced for higher momentum as $\Delta t \propto p^{-4/3}$.  The set of parameters has been extensively tested using a planar shock setup with different shock angles. We have also verified that the results do not change by using a smaller shock thickness or time step.%}
%\textcolor{blue}{Note: I am not exactly sure about the threshold you find, but since you have done an extensive test this is fine. But if you want to claim a "new" threshold, a specific study and detailed comparison with DSA is needed.}

As the shock propagates outwards, protons with an initial energy $E_0$ = 100 keV are continuously injected immediately upstream of the shock at a constant injection rate $Q$. 
The energy is roughly equal to what is required for the Parker transport equation to describe the acceleration process, i.e., the streaming anisotropy is small enough \citep{giacalone99}.
Previous kinetic studies have suggested that both parallel and perpendicular shocks can efficiently accelerate protons to this energy and the fraction can be on the order of 1\% of the thermal particles \citep{giacalone05b,guo2013}. We have not included a non-uniform seed particle population in the corona or pre-accelerated particles in the flare site, which is subjected to a future study.
%\textcolor{blue}{This implicitly assumes that the injected particles are from the background density that has a $R^{-2}$ dependence. (for this run, I didn't consider radial dependence, so at a constant rate at any location at any time)}.
A total of 1.2 $\times 10^{6}$ particles are injected.
We also use particle splitting technique to improve the statistics in the high energy portion of the particle distributions. 
Since we will only focus on the effect of streamer-like magnetic field in the low corona, the simulation is terminated when the shock propagates to 3 $R_\odot$.
We only set a free escape boundary condition along the $z$-axis at $x$ = 0, 
so a particle will be removed from the simulation if it reaches the $z$-axis or hits the solar surface.

\section{Simulation Results} \label{sec:results}

Figure 2 shows the temporal evolution of energy spectra of accelerated particles integrated over the whole simulation domain.
The vertical axis is the differential intensity $dJ/dE$, which is related to the particle distribution function by $dJ/dE = p^2f(p)$.
For a shock compression ratio $X$ = 3, the DSA theory predicts a power-law spectrum, $f(p) \propto p^{-3X/(X-1)} = p^{-4.5}$, corresponding to $dJ/dE \sim E^{-1.25}$ for non-relativistic particles.
A power-law relation with a slope of -1.25 is indicated by a black dashed line in each panel.
Note that the shock heights marked in the figure refer to that of the outermost shock front at the equator, which is 1.3 $R_\odot$ in the beginning of the simulation (at $t$ = 0).

Figure 2(a) shows the particle spectra when the shock reaches 1.5 $R_\odot$ ($t$ = 70 s), 2 $R_\odot$ ($t$ = 245 s), 2.5 $R_\odot$ ($t$ = 420 s), and 3 $R_\odot$ ($t$ = 595 s) in a streamer-like magnetic field.
At low energies the particle spectra agree well with the predicted power-law slope by the DSA theory.
At 2 $R_\odot$ and above, the particle spectra are approximately power-law distributions that roll over at $\sim$100 MeV, and the maximum particle energy is several hundred MeV.
%For example, as shown by the blue-dashed line, the particle spectrum at 2 $R_\odot$ can be well fitted by a power-law with exponential rollover at high energies, i.e., $dJ/dE \sim E^{-1.25} exp(-E/E_c)$, with the break energy $E_c =$ 70 MeV.
This indicates that particles can be sufficiently accelerated to $>$100 MeV within 2 $R_\odot$, consistent with the observations in GLEs 
\citep[e.g.,][]{reames09a,reames09b,gopalswamy12}. 
Further, after the shock passed 2.5 $R_\odot$, i.e., the outermost height of closed magnetic field lines, very few particles can be further accelerated to $>$100 MeV.
Therefore, the high energy particles above 100 MeV are mainly accelerated below 2.5 $R_\odot$. This suggests that closed fields of the streamer may play a critical role in producing high energy particles.

To further illustrate the effect of a streamer-like magnetic field configuration on particle acceleration, we compare the results for a streamer magnetic field with that for a simple radial magnetic field.
%In the two simulations, the total numbers of injected particles are the same.
%\textcolor{red}{This is not accurate.}
In Figure 2(b), it shows the integrated particle energy spectra in both cases at 2 $R_\odot$ and 3 $R_\odot$. 
The particle spectra for a radial magnetic field are denoted by the dashed lines.
In the radial magnetic field, the power-law break energy is less than 10 MeV, in comparison to $\sim$100 MeV in the streamer magnetic field case.
At 2 (3) $R_\odot$, the intensity at 100 MeV is enhanced by about three (two) orders of magnitude in the streamer magnetic field.
Therefore, it indicates that particle acceleration in the streamer-like magnetic field is much more efficient than that in a simple radial magnetic field.

Figures 3 and 4 show the spatial distributions of accelerated particles when the shock propagates to three different heights, i.e., 1.5 $R_\odot$ ($t$ = 70 s), 2 $R_\odot$ ($t$ = 245 s), and 2.5 $R_\odot$ ($t$ = 420 s), for the streamer-like and radial magnetic fields, respectively.
In Figure 3, low energy particles (2-10 $p_0$, $<$10 MeV) are shown in the upper panels and high energy particles ($>$30 $p_0$, $>$90 MeV) are shown in the lower panels.
It shows that the distribution of low energy particles is generally uniform along the shock front, while the high energy particles are mainly concentrated near the shock nose, i.e., around the top of closed field lines.
High energy particles are trapped and preferentially accelerated around the top of closed field lines, similar to the ``hot spot'' phenomenon discussed in \citet{guo10}.
As shown in panel (d), only a few particles have been accelerated to $>$90 MeV when the shock moved to 1.5 $R_\odot$. 
These particles appear around the shock nose where the shock geometry is nearly perpendicular.  Therefore it confirms that particle acceleration is very efficient at a perpendicular shock.
In addition, as shown in panel (e), we again can see particles can be accelerated to $\sim$100 MeV at 2 $R_\odot$.

Figure 4 shows the particle distribution when the coronal shock moves through a radial magnetic field. 
Note that in the lower panels a relatively lower energy range (20-30 $p_0$, 40-90 MeV) is used to show the distribution of high energy particles, because very few particles can be accelerated to that high energy ($\sim$100 MeV) as in the streamer magnetic field case (see Figure 2).
Clearly the distribution looks quite different from that in Figure 3.
As shown in panels (a) and (d), at 1.5 $R_\odot$, particles are first accelerated at shock flanks.
This is because the shock geometry is quasi-perpendicular and particles are rapidly accelerated in these regions.
As the shock keeps propagating and accelerates more particles, the low energy particles approach a roughly uniform distribution along the shock front similar to the streamer magnetic field case, while the high energy particles remain mainly concentrated at the shock flanks.

By comparing with a simple radial magnetic field, it has been shown that particle acceleration is much more efficient in a streamer-like magnetic field.
Clearly, the shock geometry is one of the factors causing the difference.
When the shock moves outward, the shock geometry evolves with time quite differently in the two magnetic configurations.
In a streamer-like magnetic field, it has quasi-perpendicular shock geometry both at shock flanks and around the shock nose.
However, in a radial magnetic field, a quasi-perpendicular shock geometry only exists at shock flanks at low heights.
As shown in Figures 3 and 4, the acceleration efficiency is much higher at a perpendicular shock compared to a parallel shock.

To further demonstrate the effect of magnetic configuration, we examine the pseudo-particle trajectory for the case with a streamer-like magnetic field.
Figure 5 shows the simulation results for a pseudo-particle finally accelerated to $\sim$65 $p_0$ ($\sim$400 MeV). 
The particle trajectory is plotted in panel (a) and the variation of particle momentum as a function of its $x$ position is shown in panel (b).
It can be seen that the particle roughly follows the closed magnetic field lines and is mainly accelerated below 2.5 $R_\odot$.
Therefore, it suggests that closed magnetic field lines of the streamer have a natural trapping effect on the particle which enables it to be confined around the shock and get accelerated repeatedly. 
%This result agrees with our previous works on electron acceleration at coronal shocks using test-particle simulations \citep{kong15,kong16}.

\section{Conclusions and Discussion} \label{sec:dicussion}

In this study, we present a numerical model to investigate particle acceleration at coronal shocks with different magnetic field configurations.
A kinematic approach is used for analytically describing the coronal shock and the coronal magnetic field.
Despite the simplicity of the model, here we point out some advantages about this approach.
First, it includes the variation of shock geometry both along the shock front and with the shock propagation.
Most of previous SEP models only consider a shock with a fixed shock obliquity angle \citep[see, e.g.,][]{verkhoglyadova15}.  % a review paper
Second, we consider perpendicular diffusion, which is important for particle acceleration at oblique shocks.
In addition, the shock can be well resolved because the shock thickness is much less than the particle diffusive length scale.
This is essential to get correct solution of DSA. 
The shock thickness resolved in this study is much less than the cell size of MHD simulations that include the calculation of particle acceleration in a similar way
\citep[e.g.,][]{kozarev13,schwadron15}.

We highlight the importance of streamer-like magnetic structures to the generation of very high energy particles.
Our calculations have shown that, as a coronal shock propagates through a streamer-like magnetic field, particles can be sufficiently accelerated to $>$100 MeV at heights below 3 $R_\odot$.
At 2-3 $R_\odot$, the particle spectra are approximately a power-law with rollover at $\sim$100 MeV and the maximum energy is several hundred MeV,
and the particle intensity at 100 MeV is enhanced by two to three orders of magnitude compared to that in a simple radial magnetic field.
Therefore, particle acceleration in a streamer-like magnetic field is much more efficient than in a simple radial magnetic field.
This is mainly due to perpendicular shock geometry and the natural trapping effect of closed magnetic fields, consistent with previous numerical studies \citep[e.g.,][]{sandroos06,guo10,kong15,kong16}.

Our study suggests that large-scale coronal magnetic configuration can be an important factor in producing large SEP events, which has strong observational implications.
It has been shown that in the 2012 May 17 GLE event, the first in solar cycle 24, the CME-shock originated below a streamer and the streamer magnetic configuration strongly affects the obliquity of the shock as it moves outward  \citep{rouillard16}.
To better understand the physical connection between large SEP events and CME/shock-streamer interactions, the large-scale magnetic configuration in the CME-shock initiation region should be examined for more events.

%\textcolor{red}{Adding the discussion about acceleration location.}
The large-scale magnetic field configuration also has an important effect on the acceleration region of energetic particles. Depending on the magnetic field configuration, particles can be accelerated either at the nose or franks of a shock.
In a streamer-like magnetic field, high energy particles are mainly trapped and accelerated around the top of closed magnetic field lines, i.e., the shock nose, while in a simple radial magnetic field, high energy particles are primarily accelerated at shock flanks. 
The low energy particles, however, are more or less uniformly accelerated along the shock front. 
A latest multi-spacecraft SEP observational study has shown the longitudinal spread width narrowing with increasing energy \citep{cohen17}. 
Our calculation shows that high energy particles are mainly accelerated and concentrated in a small region of the CME shock, which may provide an explanation for this observation.
Present SEP observations at 1 AU have mixed effects between acceleration and transport processes.
Upcoming missions such as Parker Solar Probe and Solar Orbiter will approach the source regions of SEPs in the corona, and may test the results of this study.

\acknowledgments

We thank Dr. Gary Zank and Dr. Randy Jokipii for useful comments. This work was supported by the National Natural Science Foundation of China (11503014 and 41331068), and the Provincial Natural Science Foundation of Shandong (ZR2014DQ001).
X.K. also acknowledges the financial support from the China Scholarship Council for his visit at LANL. 
F.G. acknowledges the support from the National Science Foundation under Grant No. 1735414 and U.S. Department of Energy Office of Science under Award No. DE-SC0018240. 
H.L. acknowledge the support by the DOE through the LDRD program at LANL. Simulations were performed with LANL institutional computing.

%%\appendix
%%\section{Appendix information}
\appendix
\section{Analytical solution of a streamer-like magnetic field configuration}

The coronal magnetic field is described by a magnetic flux function.
In the spherical coordinate system ($r, \theta, \varphi$), the magnetic flux function $\psi (r, \theta)$ is related to the magnetic field by 
\begin{equation}  % \[
\textbf{B} = \nabla \times \left( \frac{\psi}{r sin\theta} \hat{\varphi} \right) + \textbf{B}_\varphi, 
       \textbf{B}_\varphi = B_\varphi \hat{\varphi}.
 \end{equation} % \]

In this study, we consider a two-dimensional problem in the meridional plane, therefore $B_\varphi$ = 0.  
The other components of the magnetic field are given by:
\begin{equation}  % \[
B_r       =  \frac{1}{r^2 sin\theta} \frac{\partial \psi}{\partial \theta}, 
B_\theta   =  - \frac{1}{r sin\theta} \frac{\partial \psi}{\partial r}.
 \end{equation} % \]

 The magnetic flux function of a streamer-like magnetic field was derived by \citet{low86} as follows:
\begin{equation}  % \[
\psi_a (r,\theta) = r (1 - v^2) \left[(1 + u^2) arctan \left( \frac{1}{u} \right) - u \right] 
                  - \frac{\pi a^2 sin^2\theta}{2r}  + 2a\eta,  
 \end{equation} % \]

where

\begin{equation}  % \[
u^2 = \frac{1}{2} \left[ \left(1- \frac{a^2}{r^2} \right)^2 + \frac{4a^2}{r^2} cos^2 \theta \right]^{1/2}  
         - \frac{1}{2} \left(1-\frac{a^2}{r^2} \right),      
 \end{equation} % \]

\begin{equation}  % \[
v^2 = \frac{1}{2} \left[ \left(1- \frac{a^2}{r^2} \right)^2 + \frac{4a^2}{r^2} cos^2 \theta \right]^{1/2}  
         + \frac{1}{2} \left(1-\frac{a^2}{r^2} \right),       
 \end{equation} % \]

\begin{equation}  % \[
\eta^2 = \frac{1}{2} \left[ \left(1- \frac{r^2}{a^2} \right)^2 + \frac{4r^2}{a^2} cos^2 \theta \right]^{1/2}  
         + \frac{1}{2} \left(1-\frac{r^2}{a^2} \right),      
 \end{equation} % \]
 and $a$ is the height of the neutral point above which is the current sheet.

To have $\psi(r,0)$ = 0, $\psi(1,\pi/2)$ = 1, we can normalize the magnetic flux function by
\begin{equation}  % \[
\psi (r,\theta) = \frac{\psi_a (r,\theta) - \psi_a (1,0)}{\psi_a (1,\pi/2) - \psi_a (1,0)} .   
 \end{equation} % \]


\begin{figure}%[\htb]
\includegraphics[width=0.98\textwidth,clip,trim=0cm 0cm 0cm 10cm]{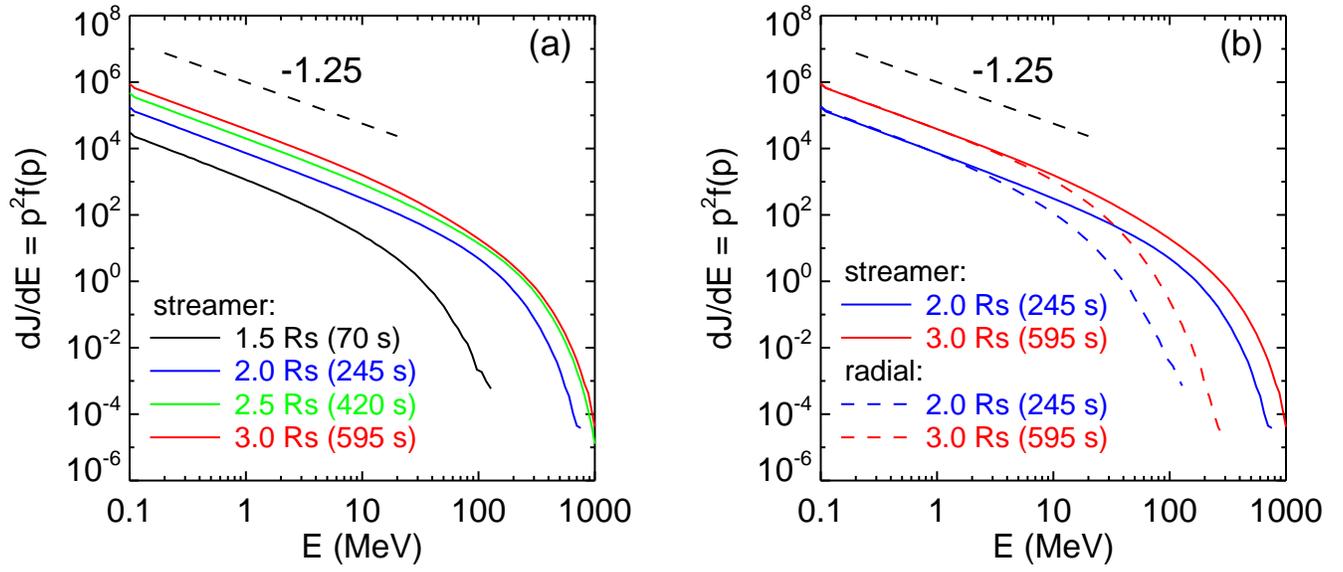}
\caption{The differential intensity of accelerated particles versus the proton kinetic energy as the shock propagates to different heights. (a) Particle spectra for the steamer-like coronal magnetic field when the shock reaches 1.5 $R_\odot$, 2 $R_\odot$, 2.5 $R_\odot$, and 3 $R_\odot$. (b) Particle spectra at 2 $R_\odot$ and 3 $R_\odot$, with that in a radial magnetic field shown by the dashed lines. The black dashed line in each panel denotes the power-law predicted by the DSA for shock compression ratio $X$ = 3.
}\label{fig:Fig2}
\end{figure}

\begin{figure}%[\htb]
\includegraphics[width=0.95\textwidth,clip,trim=0cm 0cm 0cm 0cm]{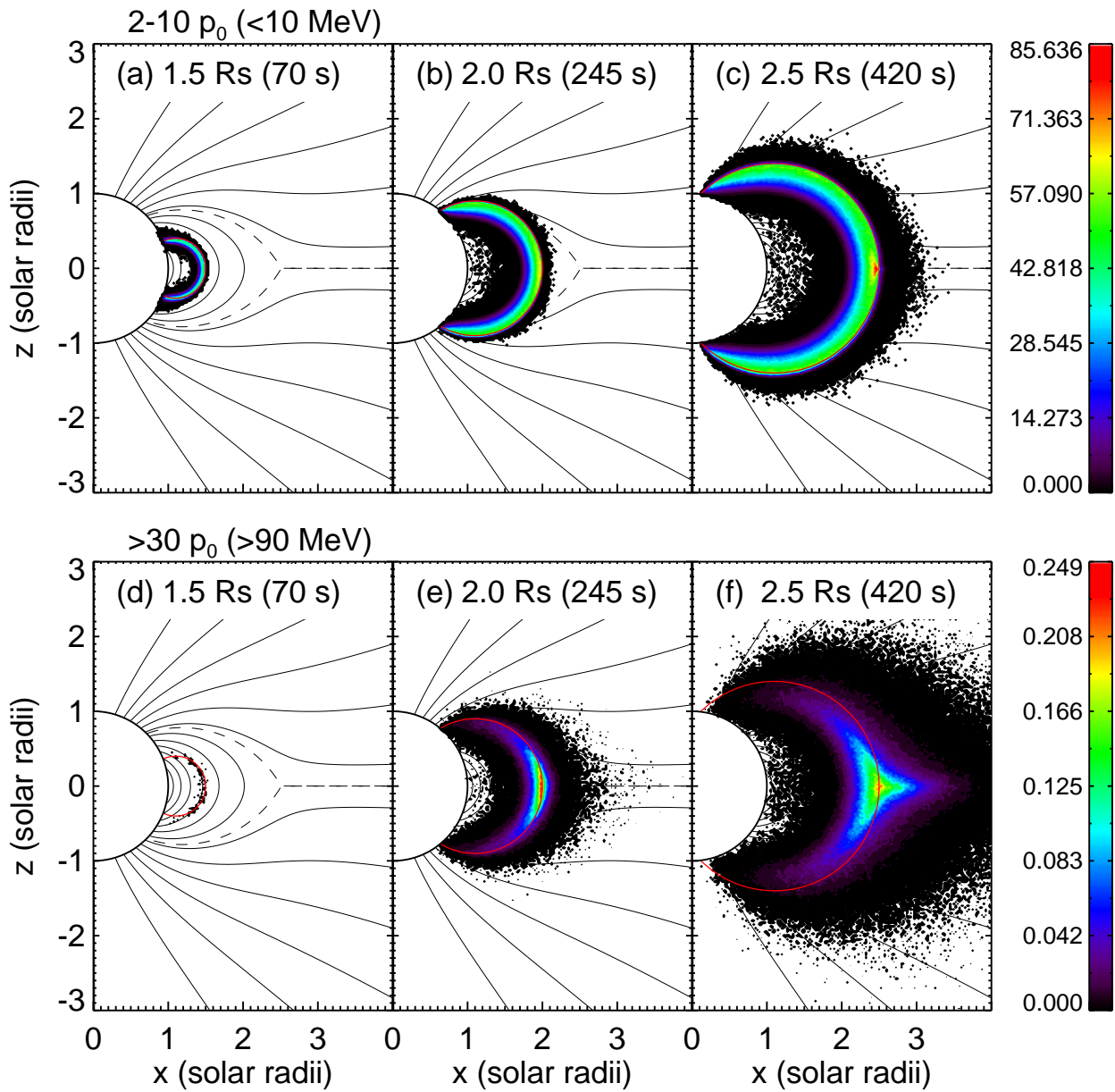}
\caption{Spatial distribution of accelerated particles as the shock propagates to three different heights, i.e., 1.5 $R_\odot$, 2.0 $R_\odot$, and 2.5 $R_\odot$. 
Low energy particles (2-10 $p_0$, $<$10 MeV) are shown in the upper panels and high energy particles ($>$30 $p_0$, $>$90 MeV) are shown in the lower panels.
The shock position is denoted by the red circle in each panel. 
}\label{fig:Fig3}
\end{figure}

\begin{figure}%[\htb]
\includegraphics[width=0.95\textwidth,clip,trim=0cm 0cm 0cm 0cm]{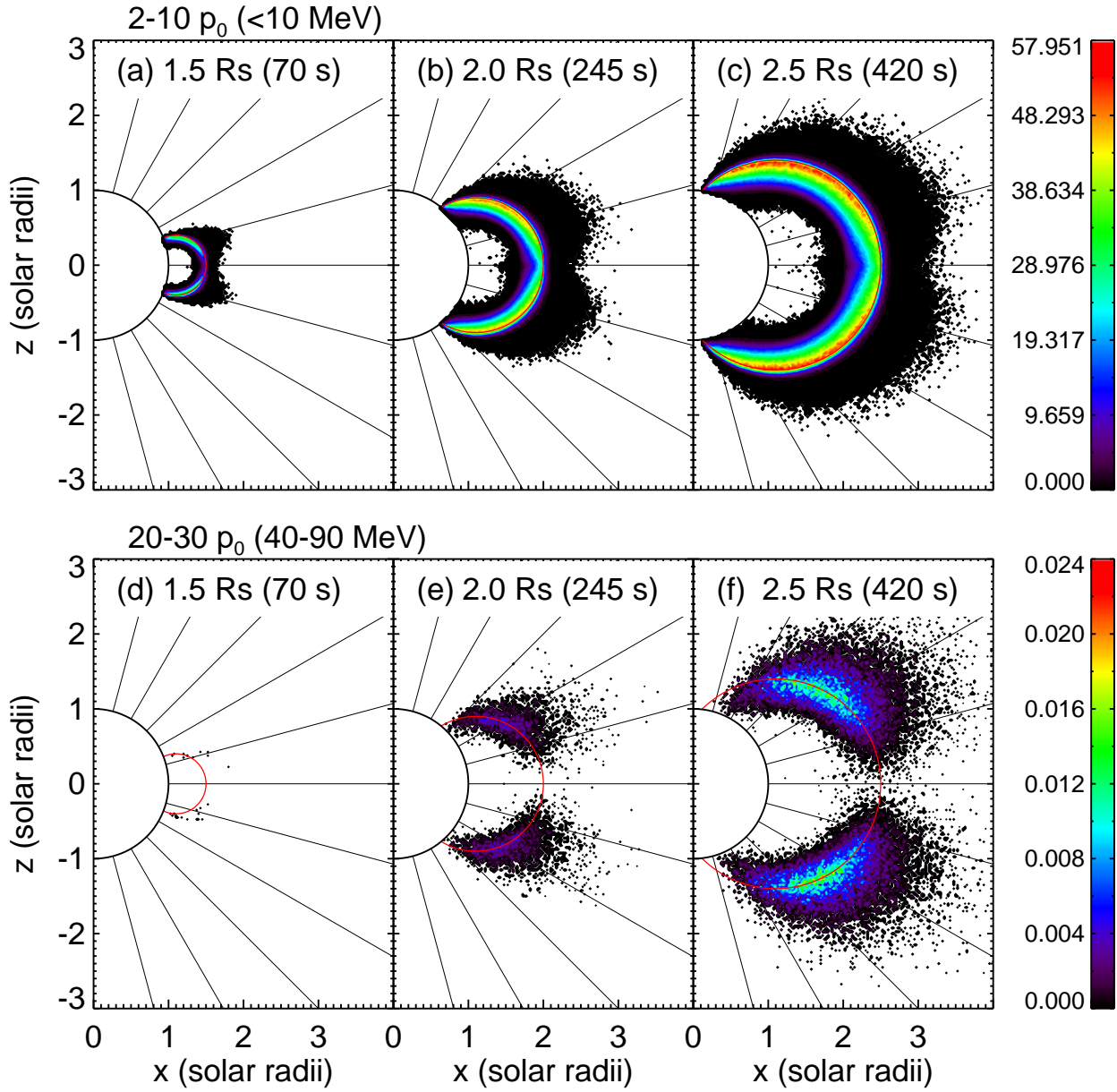}
\caption{Same to Figure 3, but plotted for the radial magnetic field case.
}\label{fig:Fig4}
\end{figure}

\begin{figure}%[\htb]
\includegraphics[width=0.95\textwidth,clip,trim=0cm 10cm 0cm 0cm]{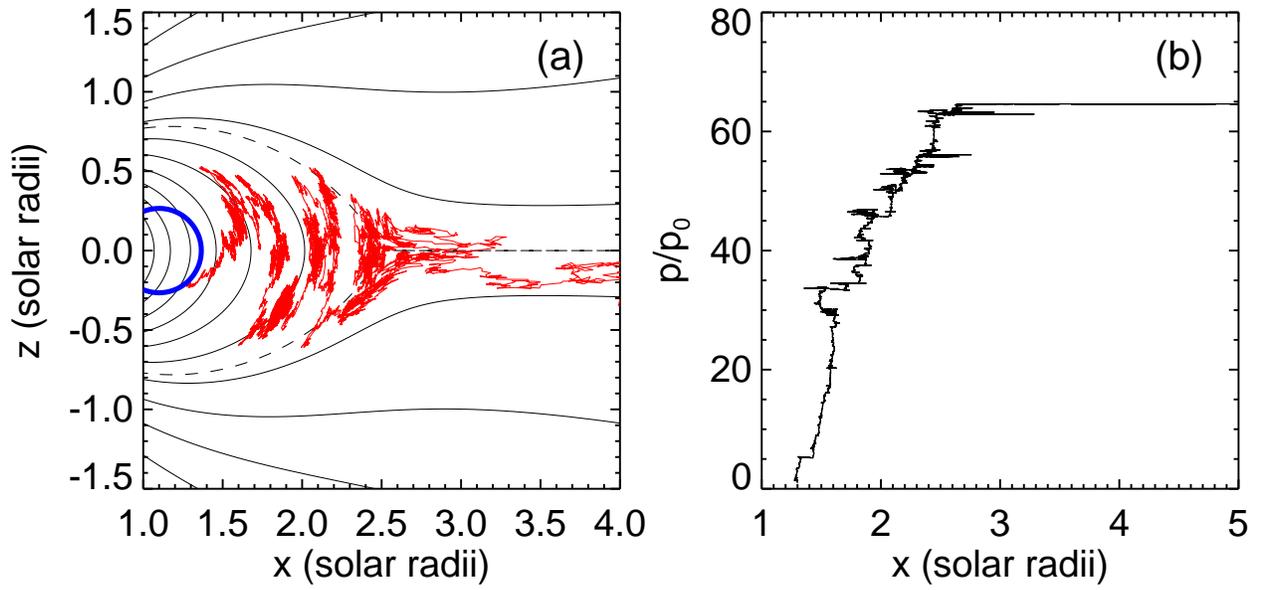}
\caption{(a) Red curve plots the pseudo-particle trajectory and the blue circle denotes the shock position when the particle is injected. (b) Temporal variation of the particle momentum as a function of its $x$ position. 
}\label{fig:Fig5}
\end{figure}

%%%%%%%%%%%%%%%%%%%%%%%%%%%%%%%%%%%%%%%%%%%%%%%%%%%%%%%%%%%%%%%%%%%%%%%%%

\end{document}